\documentstyle[12pt,aaspp4]{article}

\def\etal{{\it et~al.~}}

\def\Sec{\rlap{$^{\prime\prime}$}.\hbox to 2pt{}}
\def\Min{\rlap{$^\prime$}.\hbox to 1pt{}}

\def\Msol{\mbox{M$_\odot$}}

\begin{document}

\title{The Main Sequence Luminosity Function of Palomar 5 from
HST\footnote[1]{Based on observations with the NASA/ESA Hubble Space
Telescope, obtained at the Space Telescope Science Institute, which is
operated by AURA, Inc., under NASA contract NAS 5-26555.}}

\author{Carl J. Grillmair\footnote[2]{SIRTF Science Center, California
Institute of Technology, 1200 E. California Blvd., Pasadena, CA 91125}}

\author{Graeme H. Smith\footnote[3]{UCO/Lick Observatory, University of
California, Santa Cruz, CA 95064}}

\begin{abstract}
A low mass, large core radius, low central concentration, and strong
tidal tails suggest that the globular cluster Palomar 5 has lost a
large fraction of its initial mass over time.  If the dynamical
evolution of Palomar 5 has been dominated by the effects of mass loss,
then theoretical arguments suggest that the luminosity function should
be deficient in low-mass stars. Using deep WFPC2 F555W and F814W
photometry, we determine the main sequence luminosity functions both
near the cluster center and in a field near the half-light radius.  A
comparison of these luminosity functions yields no compelling evidence
of mass segregation within the cluster, in accord with expectations
for low-concentration clusters. On the other hand, a comparison of
the global mass function of Palomar~5 with that of $\omega$ Cen and
M55 indicates an increasing deficiency of stars with progressively
lower masses. A fit of the observed luminosity function to theoretical
models indicates a mass function for Palomar~5 of $dN/dm \propto
m^{-0.5}$, which is notably more deficient in low-mass stars than
other globular clusters that have been studied with HST. The flatness
of the mass function is consistent with models of the dynamical
evolution of globular clusters that have lost $\sim$ 90\% of their
original stellar mass.  We suggest that, like NGC~6712, Pal~5 has lost
a large percentage of its original stellar content as a result of
tidal shocking.

\end{abstract}

\keywords{Galaxy: globular clusters, Galaxy: globular clusters:
individual: Pal 5, Galaxy: evolution, Galaxy: halo}

\section{Introduction}

The halo of the Milky Way galaxy contains a number of globular
clusters that have both very low mass and low central mass
concentration. Included among such objects are the Palomar clusters
identified by Abell (1955).  Whereas much effort has been put into
understanding the internal dynamics and evolution of more highly
condensed systems, such as core-collapse clusters, the history of the
lowest concentration globular clusters ($c = \log r_t/r_c < 1.0$,
where $r_t$ and $r_c$ are the tidal and core radii) remains
uncertain. The models of Chernoff \& Weinberg (1990) suggest that
low-$c$ systems experience considerable dynamical evolution and that
they are very susceptible to tidal disruption. Consequently, clusters
that presently have low concentration are considered to have evolved
from higher concentration objects by the loss of a large fraction of
their original mass.

Suggestive evidence that the Palomar-like clusters have lost a
substantial fraction of their original mass comes from the finding by
van den Bergh \& Morbey (1984) that a number of these systems in the
outer halo exhibit an anticorrelation between integrated magnitude
$M_V$ and half-mass radius $r_h$. Smith (1985) noted that this group
of clusters follow a mass-radius relation of the form $M r_h \sim {\rm
constant}.$ Such a relation would result if these clusters originally
had similar masses and sizes, but lost different fractions of their
original mass and expanded adiabatically as a consequence, according
to the precepts of Hills (1980).

Pryor \etal (1991) argued that low-$c$ clusters lost much of their
original stellar mass over extended periods of time through the
evaporation and stripping of stars. Such star loss can be induced by
an internal process such as two-body relaxation, and by the tidal
shocking that occurs when clusters pass through the Galactic bulge or
disk (Gnedin \& Ostriker 1997; Gnedin, Lee, \& Ostriker 1999).  The
effects of dynamical relaxation on the stellar mass function with
respect to position within a globular cluster have been modeled by
Pryor, Smith, \& McClure (1986) and others. These models show that the
slope of the mass function becomes flatter in the inner regions of a
cluster as low-mass stars migrate outwards from the core.  Any
mechanism which removes stars from the outer regions of a
cluster will consequently contribute to a preferential loss of low-mass
stars. For example, the rate of evaporation of stars from a cluster
due to two-body relaxation is greater for lower-mass stars (Spitzer
1987, Giersz \& Heggie 1997). The models of Vesperini \& Heggie (1997)
show quantitatively the degree to which the stellar mass function of a
globular cluster can be altered by such processes.

Pryor \etal (1991) suggested that observational verification of such a
cluster mass-loss scenario could come from the main-sequence
luminosity function, which should show a depletion in low-mass
stars. They fitted the surface brightness and radial velocity profiles
of the cluster NGC 5466 ($c = 1.3$) with multi-mass King-Michie models
having a sharp cut-off at the lower end of the mass function. Their
best model fit to the NGC~5466 data had a global $M/L_V$ ratio of 1.0
and required a lower limit to the stellar mass function of $\sim$ 0.4
\Msol.  Since the Palomar clusters have even lower masses and central
concentrations than NGC 5466, such systems ought to provide a valuable
testing ground for these models.

Palomar~5 is one of the most accessible examples of the low-mass,
low-$c$ globular clusters in the Galactic halo. The deepest
color-magnitude diagram (CMD) of Pal~5 is that of Smith \etal (1986),
which is based on CTIO 4-m CCD data and extends to $V \sim 23.5$. The
apparent distance modulus derived by these authors is $(m-M)_V =
16.9$, so that their CMD reaches to $M_V = +6.6$, or about 2.8 mag
below the main-sequence turnoff. Below $V = 22.8$ the MSLF derived
from these data shows a deficiency of stars relative to that of
M3. However, the Smith \etal (1986) data are not definitive on this
point, since the turnover is only seen within one magnitude of the
limit of their data, and so is subject to photometric errors and
incompleteness uncertainties. To investigate the MSLF of Pal 5 to
fainter limits we present in this paper the results of deep imaging
obtained with the Wide Field/Planetary Camera 2 (WFPC2) on the Hubble
Space Telescope (HST).

Our work is underscored by the recent discovery by Leon, Meylan, \&
Combes (2000), and Odenkirchen \etal (2001) (the latter using SLOAN
Digital Sky Survey commissioning data) of strong, extensive tidal
tails emanating from either side of Pal~5. This greatly strengthens
the case for continued evaporation and stripping of stars, though
it does not rule out expansion of the cluster during its formative
period as a consequence of gas loss.

\section{Observations and Photometry}

The WFPC2 images for this program were taken on 10 and 19 August
1999. The observations are summarized in Table~1.  The two WFPC2
fields of view are shown overlaid on the Digital Sky Survey in
Figure~1.  The two fields were selected to sample both the core region
and the half-mass radius of the cluster (henceforth referred to as the
core and off-center fields). The core, half-mass, and tidal radii of
Pal~5 are 2\Min90, 2\Min96, and 15\Min9 respectively (Harris
1996),\footnote{The radii quoted are from the 1999 version of this
catalog available at the web site
http://physun.physics.mcmaster.ca/Globular.html.} giving the cluster a
central concentration of $c=0.74$.  One field was placed on the
cluster center at RA(2000) = 15:16:04.1, dec(2000) = $-$00:06:45,
while the second field was situated 2$^{\prime}$ 20$^{\prime\prime}$
south of the cluster center.  The off-center field is located just
within the half-mass radius, and well within the tidal radius.  Given
the relatively low surface density of stars in Pal 5, we opted not to
stray too far from the cluster center to be certain that we would
detect a sufficient number of stars to produce a statistically robust
luminosity function. Two 1200s and two 1300s exposures were taken in
the F814W ($I$) bandpass, while the F555W ($V$) images comprise four
1300s exposures in each field.

\begin{figure}
\epsfxsize=3.0in
\epsfbox[-280 120 265 669]{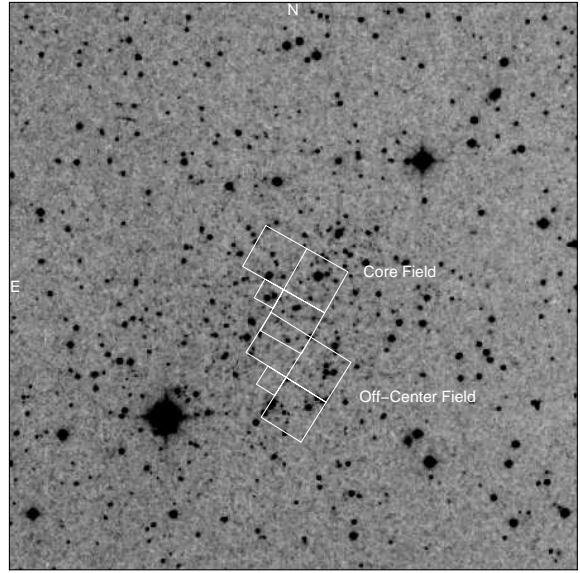}
\caption{Digitized Sky Survey image of the field containing
Palomar 5, with the WFPC2 fields examined here shown superposed.
The image shown subtends $30^{\prime}$ on a side.}
\end{figure}

\begin{deluxetable}{lcccccc}

\tablenum{1}
\tablecaption{Summary of Observations}

\tablehead{
\colhead{Field}     &
\colhead{RA(2000)}  &
\colhead{Dec(2000)} &
\colhead{Filter}         &
\colhead{Exp time (s)}  &
\colhead{Date}      &
\colhead{Dataset}
}
\startdata
Pal 5 center   &  15 16 04.18   &  $-$00 06 45.4  &  F814W  &  1200  &  10 Aug 1999  &  U4ZL0101R \nl
Pal 5 center   &  15 16 04.18   &  $-$00 06 45.4  &  F814W  &  1200  &  10 Aug 1999  &  U4ZL0102R \nl
Pal 5 center   &  15 16 04.14   &  $-$00 06 45.2  &  F814W  &  1300  &  10 Aug 1999  &  U4ZL0103R \nl
Pal 5 center   &  15 16 04.14   &  $-$00 06 45.2  &  F814W  &  1300  &  10 Aug 1999  &  U4ZL0104R \nl
Pal 5 center   &  15 16 04.18   &  $-$00 06 45.4  &  F555W  &  1300  &  10 Aug 1999  &  U4ZL0105N \nl
Pal 5 center   &  15 16 04.18   &  $-$00 06 45.4  &  F555W  &  1300  &  10 Aug 1999  &  U4ZL0106R \nl
Pal 5 center   &  15 16 04.14   &  $-$00 06 45.2  &  F555W  &  1300  &  10 Aug 1999  &  U4ZL0107R \nl
Pal 5 center   &  15 16 04.14   &  $-$00 06 45.2  &  F555W  &  1300  &  10 Aug 1999  &  U4ZL0108R \nl
Pal 5 off-center   &  15 16 04.19   &  $-$00 09 05.0  &  F814W  &  1200  &  19 Aug 1999  &  U4ZL0201R \nl
Pal 5 off-center   &  15 16 04.19   &  $-$00 09 05.0  &  F814W  &  1200  &  19 Aug 1999  &  U4ZL0202R \nl
Pal 5 off-center   &  15 16 04.15   &  $-$00 09 04.9  &  F814W  &  1300  &  19 Aug 1999  &  U4ZL0203R \nl
Pal 5 off-center   &  15 16 04.15   &  $-$00 09 04.9  &  F814W  &  1300  &  19 Aug 1999  &  U4ZL0204R \nl
Pal 5 off-center   &  15 16 04.19   &  $-$00 09 05.0  &  F555W  &  1300  &  19 Aug 1999  &  U4ZL0205R \nl
Pal 5 off-center   &  15 16 04.19   &  $-$00 09 05.0  &  F555W  &  1300  &  19 Aug 1999  &  U4ZL0206R \nl
Pal 5 off-center   &  15 16 04.15   &  $-$00 09 04.9  &  F555W  &  1300  &  19 Aug 1999  &  U4ZL0207R \nl
Pal 5 off-center   &  15 16 04.15   &  $-$00 09 04.9  &  F555W  &  1300  &  19 Aug 1999  &  U4ZL0208R \nl

\enddata
\end{deluxetable}

Photometry was carried out using the DAOPHOT II/ALLFRAME suite of
PSF-fitting routines (Stetson 1987; Stetson 1994). For each cluster
field, all eight WFPC2 frames were processed individually using
DAOPHOT II/ALLSTAR, and simultaneously using ALLFRAME. Small regions
of each frame surrounding obviously extended background objects were
excised from the final star list. The F555W and F814W magnitudes were
transformed to $V$ and $I$ using the prescription of Holtzman \etal
(1995).  A combined total of $\sim 2500$ stars were detected and
measured down to a limiting magnitude of $V \approx 27$ in the two
fields. The $I$ versus $V-I$ color-magnitude diagrams for the central
and off-center fields are shown in Figures~2a and 2b,
respectively. The $1\sigma$ formal errors in the derived $I$
magnitudes, $\sigma (I)$, as calculated by DAOPHOT II/ALLSTAR, are
plotted versus $I$ in Figure~3 for the stars found in both the core
and off-center fields.

\begin{figure}
\epsfxsize=3.5in
\epsfbox[-139 79 483 523]{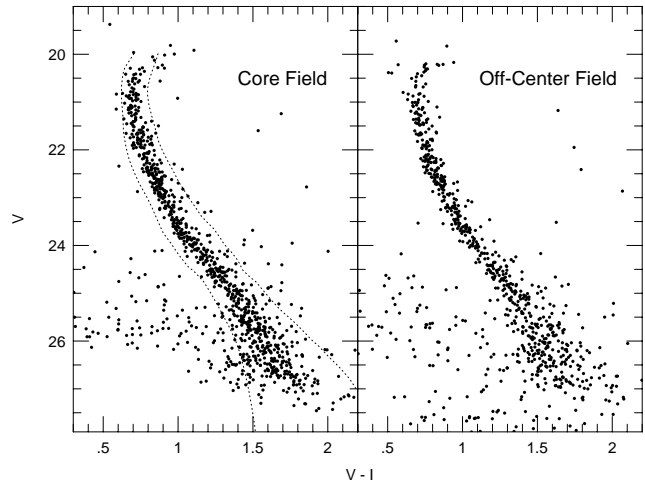}
\caption{WFPC2 color-magnitude diagrams of stars in the
core and off-center regions of Pal~5, respectively. The dashed line
shows the envelope used to count main sequence stars.}
\end{figure}

\begin{figure}
\epsfxsize=3.0in
\epsfbox[-105 50 395 500]{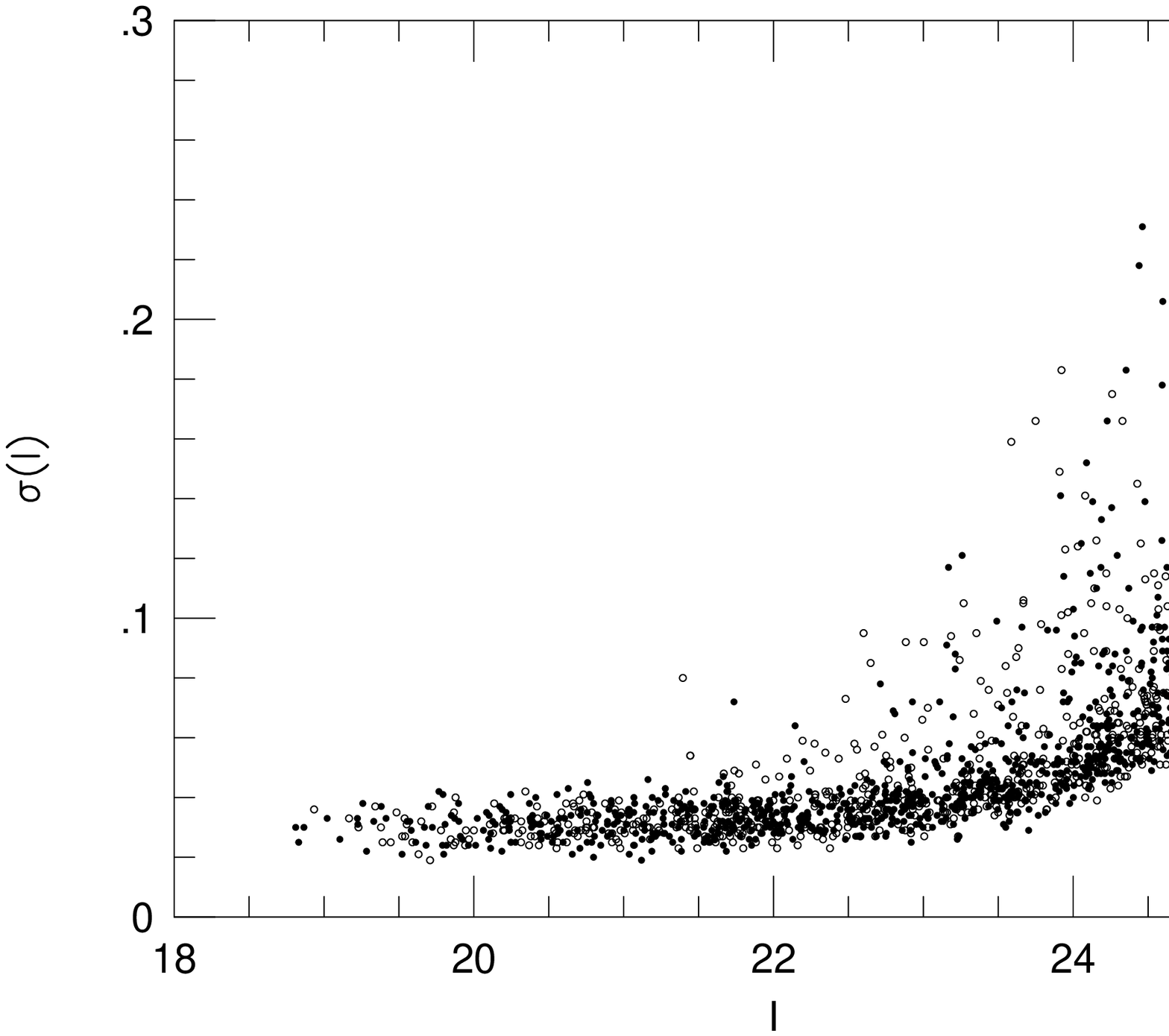}
\caption{The $1\sigma$ errors in the $I$ band magnitudes,
as calculated by DAOPHOT/ALLSTAR, are shown plotted against $I$ for
stars in the Pal~5 core field (filled circles) and off-center field
(open circles).}
\end{figure}

Completeness corrections were determined by adding to the original
WFPC2 frames 300 artificial stars at each of nine different $V$
magnitudes from $V = 23$ to $V = 27.5$. The colors of the artificial
stars were selected to follow the locus of main sequence stars, and
were identical for the core and off-center fields. The frames were
then processed using DAOPHOT II and ALLFRAME in a manner identical to
that applied to the original data (including the exclusion of regions
surrounding extended objects).  The resulting completeness fractions
are plotted in Figure~4. Owing to the very low central surface density
of stars in this cluster, the completeness fractions in the core and
off-center fields behave almost identically. Interestingly, the
largest deviations from the mean completeness curve occur, for both
core and halo fields, in the WF3 chip, which shows systematically
lower completeness fractions at all magnitudes. However, the
difference amounts to only 0.2 magnitudes, and while we correct our
luminosity functions using the completeness fractions appropriate to
each detector, we have not investigated the source of this difference
any further. Our limiting magnitude, taken to be where the
completeness fraction drops below 0.5, occurs just brightward of $V =
27$ ($I \approx 25.2$).

\begin{figure}
\epsfxsize=3.0in
\epsfbox[-200 65 370 550]{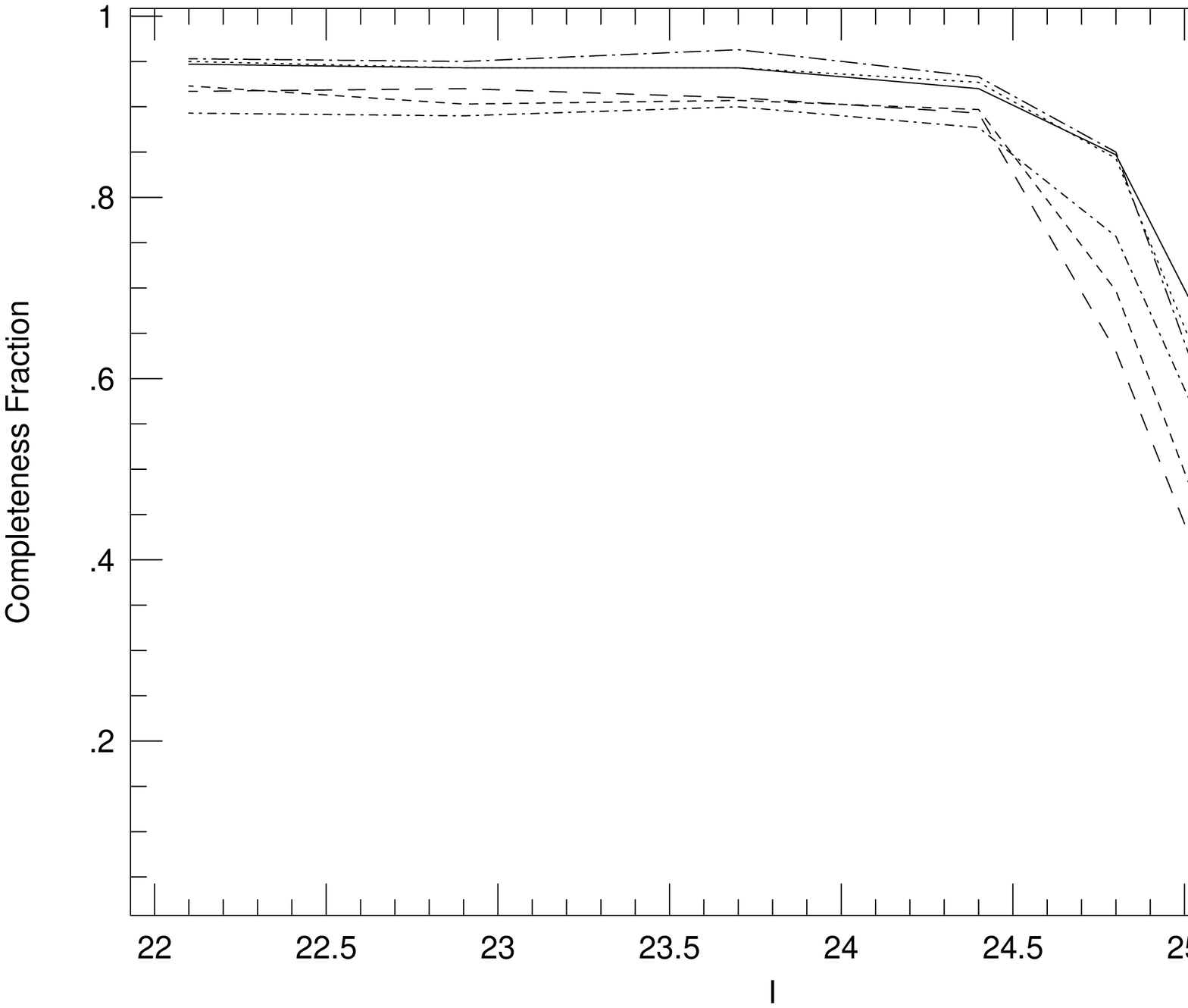}
\caption{Completeness fractions computed for both the core and
off-center fields.}
\end{figure}

\section{The Color-Magnitude Diagram and Luminosity Function}

The main sequences shown in Figures~2a and 2b are well sampled in both
fields and appear in most respects to be identical.  There are a
number of faint blue objects with $I > 23$ and $V-I < 1.2$ in both the
central and off-center fields. There are comparable numbers of such
objects in both the core and off-center fields.  Visual examination of
the WFPC2 frames reveals extended galaxies in both fields, with a
somewhat higher surface density of obviously extended objects in the
core field. On the other hand, a large number of the blue objects in
the off-center field are so faint as to prevent unambiguous
identification as background galaxies. Since the crowding and
completeness in the two fields is almost identical, it is possible
that we are seeing variations in the mean color and morphology of
background galaxies over the region spanned by the two fields.

In order to avoid contaminating our sample with blue background
objects and to measure only the stellar luminosity function (LF) of
Pal~5, we adopted the hand-drawn, main sequence envelope shown in
Figure~2.  All objects within this envelope are considered to be main
sequence stars.  For the region of the CMD within this envelope and to
$V_{lim} = 27.6$ (the limit of our completeness computations) the
observed luminosity functions in the $V$ and $I$ bands are given in
Table~2 for both the central and off-center fields, together with
corrections for completeness.  The completeness-corrected luminosity
functions are shown in Figure~5. We have transformed to absolute
magnitudes using a distance to Pal 5 of 23.2 kpc and E(B-V) = 0.03
(from the McMaster globular cluster database).  Also shown is the main
sequence LF (MSLF) of $\omega$ Cen ($c = 1.24$), as derived from
extensive WFPC2 observations by De Marchi (1999), and the MSLF of M55
($c = 0.76$) given by Paresce \& De Marchi (2000).  Of the globular
clusters for which MSLFs have been observed using HST, these are among
the most relevant for comparison with Pal 5 because of their
relatively low central mass concentrations.

\begin{figure}
\epsfxsize=3.5in
\epsfbox[-244 63 460 541]{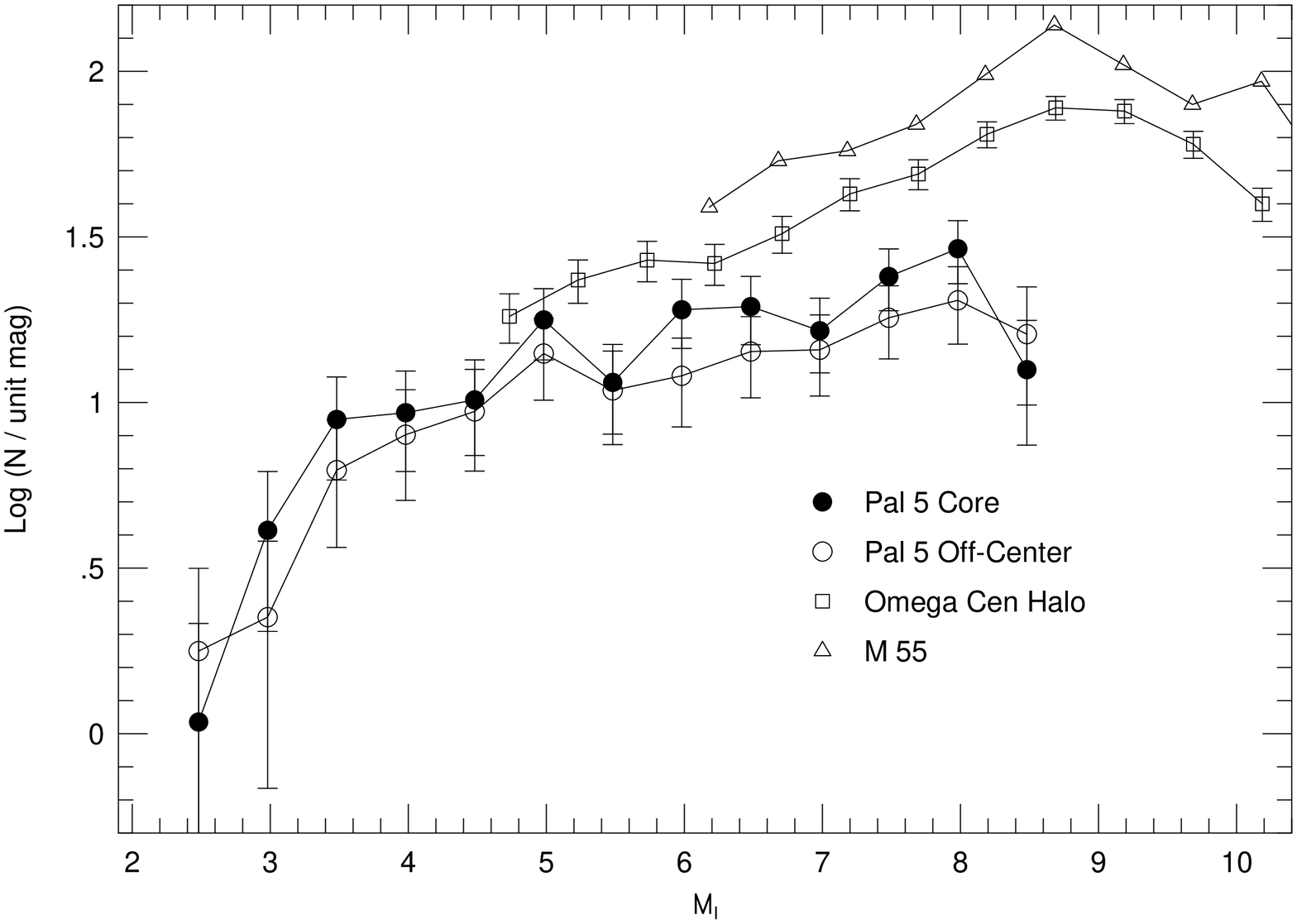}
\caption{The completeness-corrected luminosity functions
of the Palomar 5 fields, as well as the HST-derived luminosity
functions of $\omega$ Cen at $r = 4\Min6$ (De Marchi 1999) and M 55
(Paresce \& De Marchi 2000). The latter have been offset vertically
for clarity. The error bars for the Pal~5 LFs reflect statistical
uncertainties only, and do not account for uncertainties in the
completeness corrections.}
\end{figure}

\begin{deluxetable}{crrrr}

\tablenum{2(a)}
\tablecaption{Palomar 5 Main Sequence Luminosity Function - $V$ band}

\tablehead{
\colhead{$V$}    &
\colhead{N\tablenotemark{a}}      &
\colhead{N$_C$\tablenotemark{b}}  &
\colhead{N\tablenotemark{a}}      &
\colhead{N$_C$\tablenotemark{b}}  
\nl
\multicolumn{1}{c}{}      &
\multicolumn{2}{c}{Core}  &
\multicolumn{2}{c}{Off-Center}  
}
\startdata
20.00   &   2   &   2.1  &   4  &   4.3  \nl
20.50   &  20   &  21.0  &  14  &  15.2  \nl
21.00   &  39   &  40.9  &  24  &  26.0  \nl
21.50   &  36   &  37.7  &  33  &  35.3  \nl
22.00   &  41   &  43.0  &  37  &  40.2  \nl
22.50   &  63   &  66.0  &  57  &  61.6  \nl
23.00   &  58   &  60.9  &  37  &  39.8  \nl
23.50   &  47   &  49.5  &  47  &  50.6  \nl
24.00   &  68   &  72.2  &  34  &  36.7  \nl
24.50   &  61   &  64.7  &  51  &  54.9   \nl
25.00   &  61   &  64.7  &  49  &  53.1  \nl
25.50   &  78   &  82.9  &  51  &  55.5  \nl
26.00   &  85   &  92.1  &  67  &  75.7  \nl
26.50   &  87   & 112.5  &  58  &  81.5  \nl
27.00   &  33   &  65.9  &  35  &  84.9  \nl
\enddata
\end{deluxetable}
\begin{deluxetable}{crrrr}
\tablenum{2(b)}
\tablecaption{Palomar 5 Main Sequence Luminosity Function - $I$ band}

\tablehead{
\colhead{$I$}    &
\colhead{N\tablenotemark{a}}      &
\colhead{N$_C$\tablenotemark{b}}  &
\colhead{N\tablenotemark{a}}      &
\colhead{N$_C$\tablenotemark{b}}  
\nl
\multicolumn{1}{c}{}      &
\multicolumn{2}{c}{Core}  &
\multicolumn{2}{c}{Off-Center}  
}
\startdata
19.33  &   4  &   4.2  &   7  &   7.5  \nl
19.83  &  18  &  18.9  &  11  &  12.0  \nl
20.33  &  42  &  44.1  &  27  &  29.2  \nl
20.83  &  43  &  45.1  &  35  &  37.6  \nl
21.33  &  44  &  46.1  &  44  &  47.7  \nl
21.83  &  83  &  87.0  &  61  &  65.8  \nl
22.33  &  53  &  55.6  &  46  &  49.5  \nl
22.83  &  88  &  93.3  &  58  &  62.6  \nl
23.33  &  87  &  92.1  &  64  &  68.9  \nl
23.83  &  72  &  76.6  &  60  &  65.5  \nl
24.33  & 112  & 120.4  &  81  &  89.9  \nl
24.83  & 104  & 137.7  &  70  &  98.3  \nl
25.33  &  32  &  67.4  &  34  &  80.9  \nl
\enddata

\tablenotetext{a}{Observed number of stars}
\tablenotetext{b}{Number counts corrected for incompletness}

\end{deluxetable}

The Pal~5 LF appears somewhat ``bumpier'' than that of $\omega$ Cen,
and although it is tempting to attribute this to counting limitations for
the less populous Pal~5, it is noteworthy that, fainter than $M_I
\approx 4$, the same peaks and troughs are evident in both the core
and off-center fields (though perhaps more exaggerated in the core
field), lending apparent weight to their significance.

A feature in the MSLF of Pal~5 not observed in the other clusters is
a drop in the counts at $M_I = 8.5$. How robust is this
feature? Based purely on Poisson statistics, the completeness
calculations in our faintest bin are accurate to about 20\%. Thus, we
cannot rule out a mere leveling off of the counts in the off-center
field, though the drop in the core field is more significant at about
the $4\sigma$ level. Of course, this neglects possible systematic effects
which are not properly reflected in the completeness calculations, or
in the exclusion of background stars from the counts. 
It remains for deeper data to substantiate a faint turnover in the
LF of Pal~5.

Irrespective of the faintest magnitude bin in the Pal 5 WFPC2 data,
the LF slopes in the region $5.9 < M_I < 8.3$ (which is common to all
data sets) are significantly shallower for Pal~5 than they are for
$\omega$ Cen and M 55. Least-squares fits to the data in this region
yield $N \propto M_I^{0.18}$ and $N \propto M_I^{0.20}$ for M 55 and
$\omega$ Cen, respectively. For Pal~5 we obtain $N \propto M_I^{0.08
\pm 0.03}$ and $N \propto M_I^{0.11 \pm 0.04}$ for the core and
off-center fields, respectively. Thus it appears that Pal~5 is
deficient in faint main sequence stars compared to M55 and $\omega$
Cen.\footnote{This conclusion is unlikely to alter even if the
observed LF could be corrected for any background sources that fall
within the Pal~5 main-sequence region. Due to limitations on telescope
time we did not request WFPC2 images of a nearby background region
that would have enabled such a correction. Background objects might be
expected to increase in number with increasing $M_I$, so if the LF in
Fig.~5 were to be corrected for background sources, we would expect
the slope of the Pal~5 LF to be even shallower than noted in the
text.} This difference could be the result of dynamical evolution,
differences in the extent of mass segregation in these clusters, or it
might reflect a difference between the initial mass function (IMF) of
Pal 5 and the other two clusters.

Figure~6 shows the ratio in the number of stars per 0.5 magnitude bin
in the Pal~5 core field to the number in the off-center field. There
is no trend apparent in this ratio and we conclude that the two MSLFs
are identical to within the uncertainties. Also shown is the ratio of
the surface density of stars 
in the outer $\omega$ Cen field of De Marchi (1999)
to that of our core and off-center Pal~5 fields
combined. This ratio evidently increases progressively to the limits
of the data, consistent with the differences in LF slopes noted above,
and graphically illustrates the apparent depletion of faint stars in
Pal~5.

\begin{figure}
\epsfxsize=3.5in
\epsfbox[-254 50 471 545]{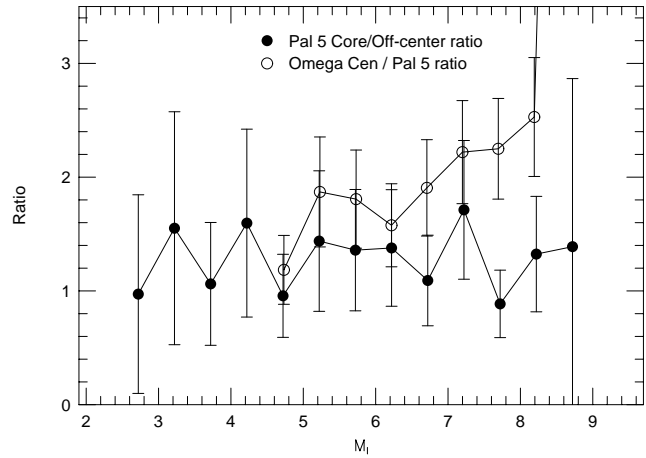}
\caption{The ratio in the number of stars per magnitude in
the core field of Pal~5 to that in the off-center field, and in the
number of stars per magnitude in the outer $\omega$ Cen field of De
Marchi (1999) to that of the core and off-center fields of Pal~5
combined.}
\end{figure}

The WFPC2 data go considerably fainter than the color-magnitude
diagram of Smith \etal (1986). Whereas their ground-based data
indicated a possible turn down in the LF at $V \sim 22.7$, this is not
seen in the WFPC2 data, which extends much fainter, and we conclude
that the former is a consequence of incompleteness and/or other
observational effects in the Smith \etal (1986) data. This further
cautions against over-interpretation of the apparent drop off in the
MSLF observed at the limit of the WFPC2 data.

\section{Discussion}

The most immediate conclusions of this paper are that the MSLF of
Pal~5 is reasonably well populated down to the limit of the
photometry, and that there is very little evidence for mass
differentiation between the core of the cluster and the half-mass
radius. This second result is in accord with the computations of Pryor
\etal (1986). Figure~1 of their paper shows that in multi-mass
King-Michie models for clusters with concentrations as low as $c =
0.7$ there is negligible variation in the local mass function within
the core radius (which for such low-$c$ clusters is close to the
half-mass radius). 
Only beyond the half-mass radius do the models of Pryor \etal
(1986) for $c = 0.7$ exhibit a steepening of the mass function due to
the migration of low-mass stars from the core to the outer regions,
but even here the effect is estimated to be small.  Within clusters
having $c < 1.0$ the Pryor \etal (1986) models therefore indicate that
the effects of mass segregation are small, and so the observed
similarity between the MSLFs in the central and half-mass region of
Pal~5 seems in accord with the models.\footnote{For clusters with $c
\geq 1.0$ mass segregation becomes more important, and the King-Michie
models show that this can cause a very measurable steepening of the
luminosity function beyond the core radius, while in the most
condensed clusters the central mass function can be depopulated to
such an extent as to cause a turn-down in the apparent mass function.}
Note however that the Pryor \etal (1986) models do not incorporate the 
effects of tidal shocking, which may be very important for Pal~5, 
as discussed below.

There is no obvious truncation to the Pal~5 MSLF, i.e., no sharp
cutoff in the mass function, in the luminosity range covered by the
WFPC2 data. The models of Pryor \etal (1991) found that a mass
function truncated sharply at $m = 0.4$ \Msol\ best fitted their
radial velocity data for the cluster NGC~5466. The WFPC2 Pal~5 data go
about 1 mag fainter than this mass limit. A relationship between
main-sequence star mass and absolute magnitude $M_V$ applicable to
Pal~5 is plotted in Figure~7 from models of Bergbusch \& VandenBerg
(1992) for stellar parameters [Fe/H] = $-1.48$, [O/Fe] = $+0.60$, $Y =
0.2354$, and an age of 16 Gyr.  This is close to the metallicity of
Pal~5. A Baraffe et al. (1997) model for [M/H] = $-1.0$, [Fe/H] = $-1.35$,
[O/Fe] = $+0.35$, and an age of 10 Gyr is also shown.
An apparent distance modulus of $(m-M)_V = 16.9$ (from the
Harris 1999 tabulation) is adopted for Pal~5.  The limit of the WFPC2
photometry at $V = 27.0$ ($M_V = +10.1$) then corresponds to a stellar
mass of 0.28 \Msol, while a mass of 0.4 \Msol\ corresponds to $M_V =
+9.2$ ($V = 26.1$). The data show no evidence for a sharp truncation
in the mass function of Pal~5 down to a stellar mass of $m \sim 0.3$
\Msol, although as noted above there is a turn-over of uncertain
significance at the faintest limits of the WFPC2 data.

\begin{figure}
\epsfxsize=3.5in
\epsfbox[-150 90 452 513]{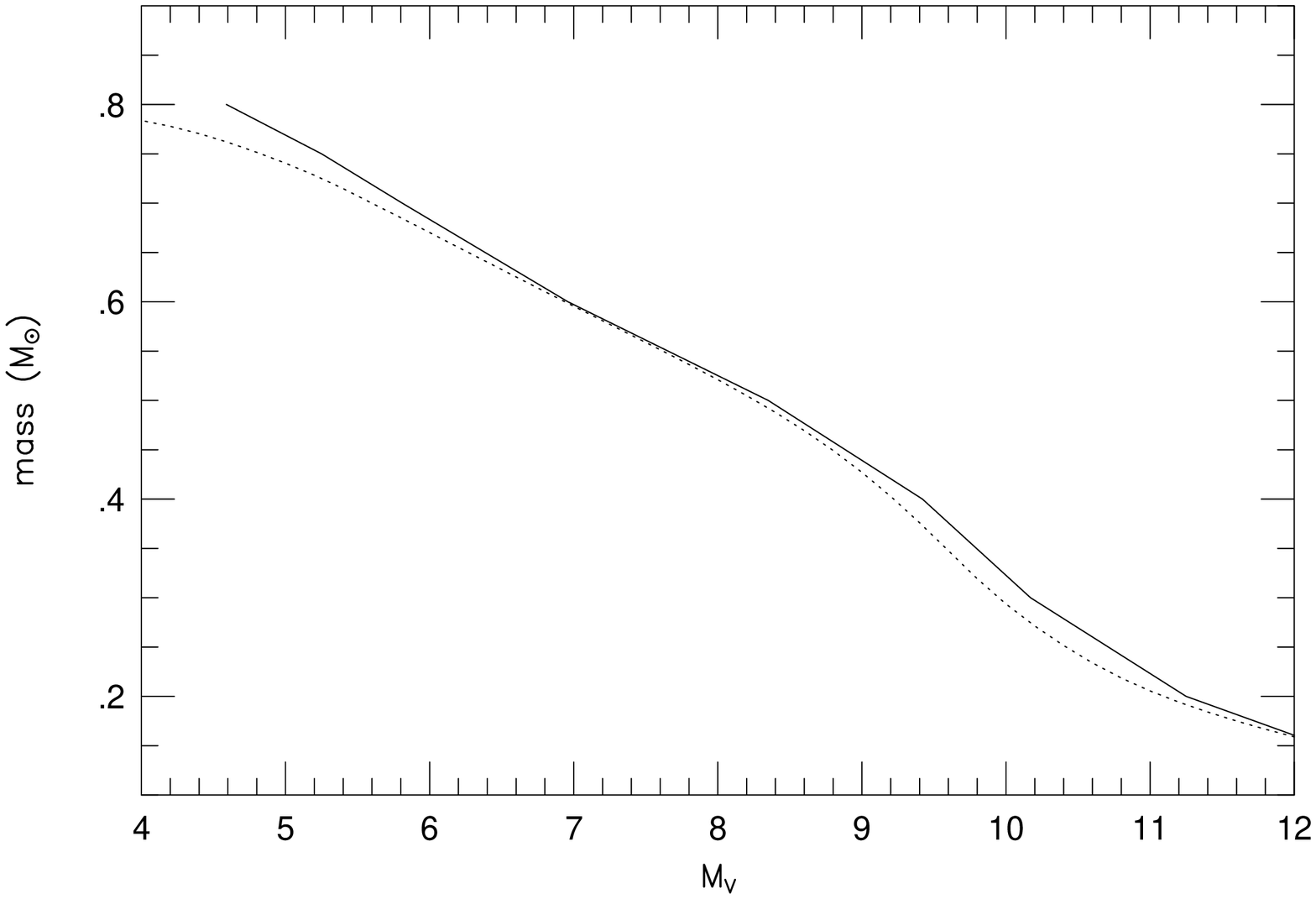}
\caption{The relation between stellar mass and absolute 
magnitude $M_V$ for 
two sets of stellar models: (i) Bergbusch \& VandenBerg (1992)
for [Fe/H] = $-1.48$, [O/Fe] = $+0.60$, $Y= 0.2354$, and an age of 16 Gyr
(shown as a dashed line), plus (ii) Baraffe et al. (1997) for [Fe/H] = $-1.35$,
[O/Fe] = $+0.35$, and an age of 10 Gyr (shown as the solid line).}
\end{figure}

The combined $V$-band luminosity function for the sum of the core plus
off-center fields in Pal~5 is shown in Figure~8. These data can be
compared to model LFs from Bergbusch \& VandenBerg (1992), which are
based on a power-law mass spectrum of the form $\phi(m) = dN/dm
\propto m^{-(1+x)}$. Their extensive grid gives values of $\log
\phi(M_V)$, where $\phi(M_V) dM_V$ is the relative number of stars in
the magnitude range $M_V$ to $M_V + dM_V$, computed for a range of
stellar compositions, ages, and values of the exponent $x$. As with
Fig.~7, their models for [Fe/H] = $-1.48$ and an age of 16 Gyr are
shown.  If their 14 Gyr models were adopted instead this would change
$\log \phi$ in the models by 0.05 at $M_V = +8.0$ for $x = -0.5$, with
smaller changes for steeper mass functions.  The model LFs 
for $x = -0.5$ and 0.5 are plotted as solid lines
in Figure~8 superimposed on the observed data.  The theoretical LFs
are normalized by Bergbusch \& VandenBerg to a value of $\log
\phi(M_V) = 1.0$ in a 0.2 mag bin centered on $M_V = +2.0$. This is
brighter than the magnitude range covered by the WFPC2, and so to
compare the models with the observational data the latter have been
normalized to a value of $\log \phi(M_V) = 2.46$ in the 0.5 mag bin
centered at $M_V = +4.6$. This normalized version of the Pal~5 LF is
related to the actual numbers of stars counted in the WFPC2 frames via
the equation $\log \phi(M_V) = \log N(M_V) + 0.6$, where $N(M_V)$ is
the completeness-corrected number of stars in the two Pal~5 fields per
0.5 mag bin.

\begin{figure}
\epsfxsize=3.5in
\epsfbox[-150 90 452 513]{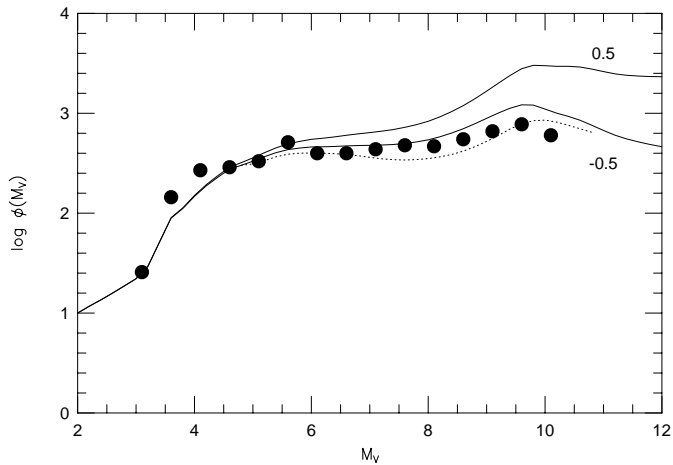}
\caption{The Pal~5 luminosity function (solid circles) for
the central and off-center WFPC2 fields combined, corrected for
incompleteness, is compared with model LFs from Bergbusch \&
VandenBerg (1992), which are shown as solid lines.  For the
observational data $\phi(M_V)$ is the relative number of stars per 0.5
mag bin normalized to match the models at $M_V = +4.6$. The models are
labeled by the value of the exponent $x$ that parameterizes the
power-law stellar mass function.  A model MSLF was also calculated for
a mass spectrum exponent of $x = -0.5$ and the Baraffe \etal (1997)
mass-magnitude relation shown in Fig.~7. This MSLF is shown as a
dashed line in Fig.~8, and is normalized at $M_V = 4.8$ to the
Bergbusch \& VandenBerg model for the same $x$ value.}
\end{figure}

We have also computed a model MSLF using the mass-$M_V$ relation of
Baraffe \etal (1997) for [M/H] = -1.0, [Fe/H] = $-1.35$, [O/Fe] =
$+0.35$, and an age of 10 Gyr.  A spline fit was made to their
relation with mass as the dependent variable, then for 0.1 magnitude
steps in $M_V$ the value of the stellar mass was computed. In the case
of the above power-law mass spectrum, the number of stars $N$ with
masses between $m_1$ and $m_2$ is $N \propto (m_2^{-x} -
m_1^{-x})$. Using this relation, the number of stars were computed per
0.2 mag bin for an exponent of $x= -0.5$ normalized at $M_V = 4.8$ to
the model of Bergbusch \& VandenBerg (1992) in Fig.~8 for the same
exponent. The resultant model LF is shown as a dashed line in
Fig.~8. Although the Baraffe \etal (1997) model gives a somewhat
different MSLF than that of Bergbusch \& VandenBerg (1992) (as one
might expect from the differences in the mass-magnitude relation in
Fig.~7), both sets of model LFs indicate that the current stellar mass
function of Pal~5 is quite flat, with a value of $x \leq -0.5$.

Piotto \& Zoccali (1999) find that the current values of $x$ for
globular clusters studied with WFPC2 range from 0.3 to $-0.5$ (where
the Salpeter mass function has a value of $x = 1.35$). The current
mass function that we derive for Pal~5 is therefore amongst the
flattest found to date in a globular cluster. Piotto \& Zoccali (1999)
showed that for the clusters in their sample the exponent $x$
correlates with the relaxation time and the cluster disruption
timescale.  One interpretation of this trend is that differences in
$x$ among globular clusters are due to differences in the degree of
dynamical evolution (Piotto, Cool, \& King 1997; Piotto \& Zoccali
1999). In this respect, the mass function of Pal~5 is in accord with
the suggestion that, at least for this cluster, the low concentration
is a consequence of the dynamical evolution and mass loss predicted by
the models of Chernoff \& Weinberg (1990).

The loss of mass, both in the form of gas expelled from individual
stars, and the escape of stars themselves from a cluster, accompanies
dynamical evolution.  Vesperini \& Heggie (1997) have modeled the
evolution of the stellar mass function of globular clusters as they
evolve dynamically and lose mass by stellar evolution, evaporation of
stars due to two-body relaxation, and the effects of Galactic disk
shocking. They find that dynamical evolution will transform an initial
stellar mass function having $x = 1.5$ into a mass function with $x =
-0.5$ by removing some 90\% of the original mass. Thus if Pal~5
started out with an initial mass function similar to the Salpeter
function these models suggest that its mass is now only $\sim 1/10$
the original mass.  The relatively low value of $x$ for Pal~5 suggests
that it has lost a larger percentage of its initial mass than the
clusters included in the study of Piotto \& Zoccali (1999).  In the
context of the models of Chernoff \& Weinberg (1990) and Vesperini \&
Heggie (1997), it seems reasonable to attribute the low overall mass
and low concentration of Pal 5 to evolutionary processes accompanied
by mass loss. The total absolute magnitude of Pal~5 is $-5.2$, which
suggests a mass of $1.1 \times 10^4$ \Msol\ for a $M/L_V$ ratio of 1.
Pal 5 may have evolved from a cluster whose initial mass was $\sim
10^5$ \Msol.

One globular cluster which provides an interesting comparison with
Pal~5 is NGC 6712. This system is more metal-rich than Pal~5 ([Fe/H] =
$-1.0$ versus $-1.4$ according to the 1999 McMaster catalog), and much
closer to the Galactic plane ($Z = -0.5$ versus 16.7 kpc for Pal~5).
The current position of NGC 6712 places it 3.5 kpc from the Galactic
Center versus 18.6 kpc for Pal~5. The $Z$-component of the space
motion of NGC 6712 was found by Cudworth (1988) to be $-122 \pm 18$
km/s, indicative of halo membership. NGC 6712 has a similar central
mass concentration ($c = 0.9$) to Pal 5, but a higher mass ($M_{V,t} =
-7.5$ versus $-5.2$), and a central luminosity density
$L_{\odot}$/pc$^3$ $\sim 8,100$ times that of Pal~5. Deep
main-sequence luminosity and mass functions have been derived from VLT
imaging by Andreuzzi et al. (2001) and De Marchi et al. (1999).
Despite a greater mass, NGC 6712 has a main sequence mass function
that is even more extreme than that of Pal~5, and is actually
inverted, so that the number of stars per unit magnitude decrease to
fainter magnitudes, even at the half-mass radius.

De Marchi et~al. (1999) attributed the depleted MSLF of NGC~6712 to
the loss of low-mass stars from the cluster as a result of tidal
stripping. As described by Gnedin \& Ostriker (1997) and Gnedin, Lee,
\& Ostriker (1999), key mechanisms which influence the rate at which
stars are lost from globular clusters are internal two-body
relaxation, and tidal shocking accompanying close passages to the
Galactic bulge and disk.  The first of these mechanisms acts in all
globular clusters, but the second is very sensitive to the orbit of a
cluster within the Galaxy. The energy input into a globular cluster
from tidal shocking due to Galactic bulge encounters is dependent on
the perigalactic distance of the cluster. The perigalacticon of
NGC~6712 is quite small, 0.9 kpc (Dinescu et~al. 1999), which lead De
Marchi et~al. (1999) to argue that bulge shocking was the main cause
of the depletion of low-mass stars from the cluster. Could such a
mechanism also be responsible for the depleted main sequence
luminosity function of Pal~5?

The orbits of both Pal~5 and NGC~6712 within the Galaxy have been
studied by Dinescu, Girard, \& van Alterna (1999) and Odenkirchen
et~al. (2001).  Using the measured space motion of Pal~5, Dinescu
et~al. (1999) calculated that this cluster is an orbit of eccentricity
$e = 0.74 \pm 0.18$ and a perigalacticon of 2.3 kpc. Thus even though
Pal~5 is currently situated at relatively large distances from the
Galactic Center and above the Galactic plane, the orbit is such that
close encounters with the Galactic bulge could be important. In fact,
Dinescu et~al. (1999) find that in the case of both Pal~5 and NGC~6712
tidal shocking has a greater influence on cluster evaporation than
does two-body relaxation. By contrast, Odenkirchen et~al. (2001) find
from modeling of the morphology of the tidal tails of Pal~5 that the
perigalactic distance is a much less extreme 7 kpc, but they do
calculate that Pal 5 has made three passages through the disk of the
Galaxy over the past 500 Myr, two of which occurred at Galactocentric
radii of 8-9 kpc. Thus disk shocking may have been a major contributor
to mass loss from Pal~5. The tidal tails contain $\sim 1/3$ the number
of stars as the cluster itself, indicating that whatever the mechanism 
responsible for these tails, it continues to remove a considerable
amount of mass from the cluster.

Despite their current differences in total mass, it is possible that
both Pal~5 and NGC~6712 have lost a large fraction of their initial
mass due to tidal shocking induced either by bulge or disk encounters.
We speculate that NGC 6712 originally began as a more massive cluster
than Pal~5, but because it has evolved in an orbit passing closer to
the Galactic bulge, it has been subjected to more extreme tidal
shocking and concomitant stripping. NGC 6712 has perhaps lost a
greater fraction of it's original mass than Pal~5, but remains more
massive than Pal~5 today simply because it had a much larger
mass to begin with.

Calculations of the disruption time of the ``current'' version of
Pal~5 have been made by both Gnedin \& Ostriker (1997) and Dinescu
et~al. (1999), though their results differ substantially. Based on the
Galactic orbit derived from the observed space motion, Dinescu
et~al. (1999) computed a disruption time of only 0.1 Gyr. Gnedin \&
Ostriker (1997) adopted a more statistical formalism for treating
cluster orbits and derived disruption times of 1 Gyr and 50 Gyr
according to two different Galactic mass models. Hence, while there
remains considerable uncertainty, these calculations are entirely
consistent with the idea that tidal shocking and stripping played a
major role in depleting mass function of Pal~5

The above arguments suggest that mass loss may have been an important
factor in determining particularly the low-mass end of the Milky Way
globular cluster mass function (GCMF). Theoretical models by Vesperini
(1998) demonstrate this point, showing that even an initial power-law
GCMF can be turned into a log-normal mass function as a consequence of
dynamical evolution and cluster mass loss. If so, flat or declining
stellar luminosity functions such as those of Pal~5 and NGC 6712
should be expected among other clusters which populate the faint end
of the Milky Way GCMF. This would include systems such as Pal~13
(Siegel \etal 2001) and E3 (van den Bergh, Demers, \& Kunkel 1980),
whose depleted giant branches, low total mass and concentration, and
high specific frequency of blue straggler stars and binaries are all
consistent with the preferential loss of lower-mass stars. The main
sequence luminosity function of E3 derived from deep ground-based
images does indeed decline towards faint magnitudes (McClure \etal
1985) in a way that recalls NGC 6712.  As our knowledge of the orbital
kinematics of Glactic globular clusters grows, understanding the
causes and effects of stellar mass loss in globular clusters may well
have important consequences for our understanding of the structure and
formation of both our Galaxy and other galaxies with globular cluster
systems. In the meantime, constraining the models will clearly benefit
from a larger sample of HST-quality LFs of the lower-mass Milky Way
globulars.

\end{document}